\documentclass{article}[12pt]
\usepackage{graphicx,cite,epsfig}
\newcommand{\eeq}{\end{equation}}
\newcommand{\barr}{\begin{array}}
\newcommand{\earr}{\end{array}}

\def\gev{\: \rm GeV} 
 
\newcommand{\dis}{\displaystyle}

\def\beq{\begin{eqnarray}}
\def\eeq{\end{eqnarray}}
\def\ba{\begin{array}}
\def\ea{\end{array}}

\def\a{\alpha}

\def\g{\gamma}
\def\l{\lambda}
\def\s{\sigma}

\def\slash{\!\!\!\!/}

\def\l{\lambda}
\def\e{\ell}
\def\snu{\widetilde{\nu}}

\def\stau{\widetilde{\tau}}

\def\slep{\widetilde{\ell}}
\def\N0{\widetilde{\chi}^0}

\def\Cpm{\widetilde{\chi}^\pm}

\def\f{\phi^{--}}
\def\h{\phi^{\pm\pm}}

\begin{document}
\begin{flushright} HIP-2007-74/TH \\ \end{flushright}
\begin{center}
{\Large Associated Photons and New Physics Signals at Linear Colliders}
\\ \vspace*{0.4in}
{\large Santosh Kumar Rai} \\ 

\vspace*{0.2in}
\centerline {(santosh.rai@helsinki.fi)}

\vspace*{0.4in}
{\sl High Energy Physics Division, Department of Physics, University
     of Helsinki,\\ and Helsinki Institute of Physics, P.O. Box 64,
     FIN-00014 University of Helsinki, Finland\\ \rm }
\end{center}
\vspace*{0.6in}
{\large\bf Abstract}

We study signals for beyond standard model physics and consider the 
virtues of single photon signals or associated photons in the final
states in identifying different scenarios of new physics models in a 
very efficient and novel way.

\vspace*{0.2in}

\vspace*{0.2in}

\vspace*{0.6in}

\section{Introduction} 

Human scientific knowledge has come a long way in its quest for solutions to
the unanswered questions and mysteries of nature, starting from the
discovery of molecules and atoms to the present day knowledge of the
smallest constituents of matter. And what we know today about the most
fundamental building blocks of matter and the nature of forces governing
their interaction is at best explained by the {\it Standard Model} (SM) of
particle physics. Despite the remarkable successes, it is
unlikely that the Standard Model is actually a complete theory of the
fundamental laws of nature. Although the standard model is a mathematically
consistent renormalizable field theory whose predictions have matched and
withstood experimental tests down to at least $10^{-16}$cm, with an
exception of the Higgs sector, it still leaves us with a lot many unresolved
theoretical issues, like for instance the origin of mass, CP violation,
number of fermion flavours, the hierarchy problems, etc.

Experimental hints for neutrino masses\cite{neuexp} already provide us with
the necessity to consider physics beyond the SM. Another important issue
with the SM is the stabilization of the electroweak scale and the origin of
mass. SM requires an elementary scalar in the theory which is responsible
for the mass of all the particles in SM through the Higgs
mechanism\cite{higgsM}.  However, the physical state associated with this
scalar, the Higgs boson, is yet to be experimentally observed. Electroweak
precision data require the mass to be less than $\sim$ few 100 GeV. Further
more the Higgs mass squared receives large quantum corrections which drive
its mass to the cut-off of the theory. One needs physics beyond the SM to
stabilize the Higgs mass which is related to the scale of electroweak
symmetry breaking in SM, in order to get the natural scale. Also the very
fact that gravity is not fundamentally unified with the other interactions
in the SM and there is no way to generate a quantum theory for gravity
within the SM leads us to explore new theoretical ideas that extend beyond
the SM and try to address physics issues concerning the SM. The nice feature
of many of the proposed models is that they are predictive and should assert
themselves at the TeV scale.

The world is sitting at this energy frontier with the running of Tevatron at 
Fermilab and the advent of the Large Hadron Collider (LHC) at CERN which would 
help us probe this disillusioned energy scale for any signal of new physics 
beyond the standard model and the origin of electroweak symmetry breaking 
(EWSB) with its associated mechanism that endows masses to the elementary 
particles. With a creative stream of tentative answers steadily flowing in and 
a distinct trait that seems common to all proposals, that some kind of new
physics phenomenology must exist at the scale of TeV, pushing the SM as some
form of an effective theory in the low energy limit we may be sitting at the
cross-roads of discovery. It is expected that candidate
theories like supersymmetry, technicolour, little Higgs and models of
extradimensions, to name a few, would be established or highly constrained.

The Large Hadron Collider (LHC) would mainly act as a discovery machine like
all other hadron colliders, and it is expected that a new $e^+ e^-$ Linear
Collider (LC)\cite{ILC} would complement the discoveries at LHC and make
possible precision measurements of the parameter space governing the new
physics scenario.  We shall focus on a few of the many promising candidates 
of beyond standard model
(BSM) physics scenarios. We look at their discovery prospects at future
linear colliders through a particular channel of production, in association
with photons in the final state. In section~\ref{sec:asps} we discuss the
nice features of the associated photon signals at the next generation linear
colliders and also try and motivate the idea to look for new physics signals
through this production mode. In section~\ref{sec:npsign} we take up
different new physics models and discuss the proposed signal in isolating
its signatures. We finally summarize and present our conclusions in
section~\ref{sec:concl}.

\section{Associated Photon Signals at Linear Colliders}
\label{sec:asps}

The history of associated photon signals to look for new physics signals
goes way back\cite{history,Fayet,Ellis,Grifols} and has been a process of great 
interest for the physics programme at LEP\cite{mont1,mont2}. 
It is also expected to be an important mode for new physics signals for future 
linear $e^+e^-$ colliders\cite{mont1}.  
The production of one or more photons in the final state along with the
electroweak gauge bosons of the SM has been used extensively to probe the
gauge interactions and the anomalous nature of photon interactions with
other gauge bosons in the SM\cite{anomalous1,anomalous2}. 
These studies have looked for new physics effects by looking at trilinear 
and quartic self-interactions of
the gauge bosons involving photons. Precise knowledge of such interaction
would help us understand the gauge structure of the underlying physics. The
most important result provided by single photon events is however the
precise measurements of the number of light neutrino types\cite{Ma,PDG} 
$N_\nu$ which is obtained by measuring the cross section of the process 
$e^+e^-\to\gamma \nu\bar{\nu}$. This process is also an important mode for new
physics searches as it is sensitive to contributions from physics beyond the
SM\cite{godfrey,choi}. 

We shall focus on the particular process
\beq
e e \to \gamma + X 
\label{eqn:process}
\eeq
where $X$ can be any weakly interacting massive particle belonging to
scenarios of new physics beyond the SM. We choose not to write the charges
on the colliding particles of the beams, as we will consider two different
cases, where the collision is either of electron-positron beams or
electron-electron beams. The particle produced in association with $\gamma$
carries two units of charge $e$ when the collision is between two electron
beams. The above process can be a very efficient tool to search for new
physics signal as we show through the examples in the next section. 
The case of ``photon+missing energy" is the most widely
studied signal, which within the framework of SM accounts for the
neutrino-counting experiment. It also gives an independent probe for the
$\gamma W^+ W^-$ coupling at the high energy colliders as the contribution
from the $W$-boson exchange becomes dominant. Thus it serves as an efficient
tool to study anomalous couplings of the photon with the $W$ boson as well.
Several studies exist in the literature which consider the ``photon+missing
energy"  process at future linear $e^+e^-$ colliders to look for new physics
signals\cite{sinphot1,sinphot2,sinphot3,sinphot4,raigrav,sinphot5}.

It is needless to say that linear colliders will have the ability to make
precise test of the structure of electroweak interactions at very short
distances. Looking at the simplest process of $e^+e^- \to f\bar{f}$, the SM
cross-section prediction can be put in the form
\beq
\frac{d\s}{d\cos\theta}&(&e^-_Le^+_R \to f_L\bar{f}_R~~) =
\frac{\pi\a^2}{2s}N_C \nonumber \\
&.&\left|Q_f + \frac{(\frac{1}{2}-\sin^2\theta_w)(I^3_f-Q_f\sin^2\theta_w)}
{\cos^2\theta_w\sin^2\theta_w}\frac{s}{s-m^2_Z}\right|^2 .(1 +
cos\theta)^2
\label{eqn:ffscatter}
\eeq
where $N_C=1$ for leptons and 3 times for quarks, $I^3_f$ is the weak
isospin of $f_L$, and $Q_f$ is the electric charge. For $f_L$ production,
the $Z$ contribution typically interferes with the photon constructively for
an $e^-_L$ beam and destructively for an $e^-_R$ beam. Thus, initial-state
polarization is a useful diagnostic at the LC. Applied to familiar
particles, they would provide a diagnostic of the electroweak exchanges that
might reveal new heavy weak bosons or other types of new interactions. 
Simple annihilation processes can also be used to test for new interactions.
However the best option to study such electroweak exchanges would be to
study their physics at its resonance. The obvious reason being that,
off-shell contributions will be strongly propagator-suppressed and suppress the
new physics signal drastically. Our motivation for considering the process 
given in Eq.~(\ref{eqn:process}) is principally based on the fact that design 
for the future linear collider allow the machine to run at one or a few fixed 
center-of-mass energies. Single photon signals will allow the on-shell 
production of a massive particle $X$ as long as $M_X < \sqrt{s}$. We show that 
this process can have a resonant production for $X$ which shows up in the 
energy distribution of the photon. The idea is that as the photon carries 
away a variable amount of energy, it is possible for the remaining system 
(assuming $X$ decays to some final states) to strike a $s$-channel resonance 
of the particle $X$, just as initial state radiation (ISR) at LEP-2 has been 
seen to cause a `radiative return« to the $Z$-boson pole. Then one can expect 
additional bump(s) over the continuum SM background in the photon energy 
distribution. The photon energy will be uniquely fixed by the well-known 
formula
\beq
E_\gamma = \frac{s - M_X^2}{2\sqrt{s}}
\label{eqn:resonance} 
\eeq 

This signal is particularly interesting because of its simplicity and
cleanliness. In the next section we discuss different physics models beyond
SM and how they can leave their imprint on the ``associated photon" signals
at the future linear colliders.
 
\section{New Physics and Associated Photon Signals}
\label{sec:npsign}
In this section we highlight the use of associated photon signals as a
search tool for resonances at future linear colliders for massive particles
predicted in various new physics models whose mass is less than the
center-of-mass energy ($\sqrt{s}$) of the machine. Such resonances are very
likely to be missed if they are produced off-shell in the s-channel, and if 
their mass is also quite less than the ($\sqrt{s}$) of the machine. We show
that an associated photon carries the mass information of the produced
particle in its energy distribution as given by Eq.~(\ref{eqn:resonance}).
We investigate different models and study their features through the
proposed signal.

\subsection{Supersymmetry}

The most extensively studied new physics scenario over the last three
decades has been supersymmetry (SUSY). If one considers $R$-parity ($R_p$)
conserving SUSY, ($R$ is defined as $R=(-)^{L+3B+2S}$, where $L,B$ and $S$
stand, respectively, for the lepton number, baryon number and spin of a
particle) then most of the search strategies are based on the fact that the
lightest supersymmetric particle (LSP) is a massive weakly interacting
neutral superparticle. It is stable and escapes detection and thus one
expects large missing energy associated with SUSY signals. The analogue to
our ``single photon" signal in SUSY would be production of a pair of LSP's
with a photon in the final states which would invariably affect the cross
section of the process 
$e^+e^- \to \gamma E\slash$ \cite{sinphot1,sinphot3,sinphot4,sinphot5}. 
However, $R$-parity can be violated\cite{RPV} as long as either lepton 
number $L$ or baryon number $B$ if not both, is conserved. This can scramble 
the SUSY signals quite dramatically as the LSP is no longer stable and can 
decay within the detector. Admitting lepton number violating operators of 
the $LL\bar{E}$ form (where we have assumed the conservation of baryon 
number $B$), the relevant term in the superpotential can be written as
\begin{equation} 
{\cal W}_{LL\bar{E}} = \l_{ijk} \epsilon_{ab} \hat{L}_i^a
\hat{L}_j^b \hat{E}_k  \ , \qquad i, j = 1\dots 3
\label{eqn:superpot}
\end{equation} 
where $\hat{L}_i \equiv (\hat{\nu}_{Li}, \hat{\ell}_{Li})^T$ and $\hat E_i$
are the $SU(2)$-doublet and singlet superfields respectively whereas
$\epsilon_{ab}$ is the unit antisymmetric tensor. Clearly, the coupling
constants $\l_{ijk}$ are antisymmetric under the exchange of the first two
indices;  the 9 such independent couplings are usually labelled keeping
$i>j$. Written in terms of the component fields, the above superpotential 
leads to the interaction Lagrangian
\beq 
\begin{array}{rcl}
{\cal L}_\l &
= & \dis \l_{ijk} \bigg[
   \snu^i     \bar{\e}_R^k           \e_L^j
+  \slep_L^j     \bar{\e}_R^k           \nu_L^i
+ (\slep_R^k)^*  \overline{(\nu_L^i)^c} \e_L^j
+ (\snu^i)^*  \bar{\e}_L^j           \e_R^k
+ (\slep_L^j)^*  \bar{\nu}_L^i         {\e}_L^k
+  \slep_R^k     \bar{\e}_L^j          (\nu_L^i)^c 
\\
& & \dis \hspace*{0.3in}
-  \snu^j     \bar{\e}_R^k           \e_L^i
-  \slep_L^i     \bar{\e}_R^k           \nu_L^j
- (\slep_R^k)^*  \overline{(\nu_L^j)^c} \e_L^i
- (\snu^j)^*  \bar{\e}_L^i           \e_R^k
- (\slep_L^i)^*  \bar{\nu}_L^j         {\e}_L^k
-  \slep_R^k     \bar{\e}_L^i          (\nu_L^j)^c
\bigg] \ .
\earr
     \label{gen_Lag}
\eeq
The $R\slash_p$ couplings are constrained by various
experiments\cite{rplimits}. However for our case, we are interested in a
process which deals with one particular type of coupling which allows single
sneutrino production in association with a hard photon.
The terms relevant for our discussion are the first and fourth ones on both
first and second lines of Eq.~(\ref{gen_Lag}), with $j = k =1$ on the first
line and $i = k = 1$ on the second. 

The sneutrino decay width, which never rises above 3--4 GeV, allows us to
apply the {\it narrow-width approximation} and therefore, we solely consider
on-shell production of sneutrinos (of muonic or tauonic flavour). 
If, indeed, $m_{\snu} < \sqrt{s}$, then the cross-section for
$e^+ ~+~ e^- \longrightarrow \snu_{\mu/\tau} (\snu_{\mu/\tau}^*)$ will be
strongly propagator-suppressed. As discussed in section~\ref{sec:asps} the 
processes of interest for us then becomes
\beq
\barr{rcl} e^+ ~+~ e^- &
\longrightarrow & \gamma ~+~ \snu_{\mu/\tau} (\snu_{\mu/\tau}^*) 
\longrightarrow
\gamma ~+~ e^+ ~+~ e^- 
 \\ 
& \hookrightarrow & \gamma ~+~ \snu_{\mu/\tau}
(\snu_{\mu/\tau}^*) \longrightarrow \gamma ~+~ \nu_{\mu/\tau}
(\overline{\nu}_{\mu/\tau}) ~+~ \N0_{1/2/3/4}
  \\
 & \hookrightarrow & \gamma ~+~ \snu_{\mu/\tau} (\snu_{\mu/\tau}^*)
 \longrightarrow \gamma ~+~ \mu^\mp/\tau^\mp ~+~ \Cpm_{1/2} \ ,
\earr
\eeq
where the application of the narrow-width approximation ensures an almost
monochromatic photon of energy given by Eq.~(\ref{eqn:resonance}), where $X$
is to be replaced by the sneutrino mass. This, potentially, would stand out
against the continuum spectrum arising from the SM background. Since the
sneutrino $\snu_{\mu/\tau}$ can have a variety of decay channels, we can
simply tag on a hard isolated photon associated with any of these decay
channels and look for a line spectrum superposed on the continuum
background. This will lead to clear signals of sneutrino production.
Moreover, the $R$-parity-violating decays of the sneutrino will set up
multi-lepton final states (with associated photons) which will have little
or no SM backgrounds worth considering. For such states a mono-energetic
photon will clinch the issue of sneutrino production\cite{raisneutrino}.
The specific reaction on which we focus in this case is the associated
photon process
$$
e^+ ~+~ e^- \longrightarrow \gamma ~+~ \snu_{\mu/\tau} (\snu_{\mu/\tau}^*) ~~.
$$ 
The squared and spin-averaged
matrix element for this is, then
\begin{equation}
\overline{|{\cal M}|^2} = 8\pi\alpha~\l_{1j1}^2 \frac{s^2 + \widetilde{m}_j^4}{tu}
~\theta(s - \widetilde{m}_j^2)
    \label{mesq}
\end{equation}
where $\widetilde{m}_j$ is the mass of the muonic ($j=2$) or tauonic
($j=3$) sneutrino. 
The collinear singularity in Eq.~(\ref{mesq}), so characteristic of massless
electrons and photons, is automatically taken care of once one imposes
restrictions on the phase space commensurate with the detector acceptances.
In the rest of the analysis, we shall require the photon to be sufficiently
hard and transverse, namely

\beq
\barr{lcl}
{\rm pseudorapidity}: & \qquad& 
    \left| \eta_\gamma \right| < \eta_\gamma^{(max)} = 2.0 \ , \\
{\rm transverse \ momentum}: & & p_{T\gamma} > p_{T\gamma}^{(min)} = 20 \gev \ .\earr
    \label{photon_cuts}
\eeq
The cross-section is plotted in Fig.~(\ref{fig:prodn}) as a function of
$\widetilde{m}_j$ and with $\l_{1j1} = 0.03$ for a linear collider running
at ($a$)~500~GeV and ($b$)~1~TeV.
\begin{figure}[th] 
\centerline{\psfig{file=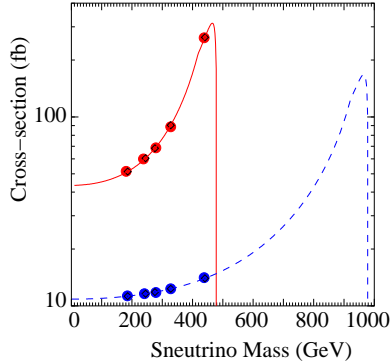,width=2.0in}}
\vspace*{8pt} 
\caption{\footnotesize\it Cross sections for
sneutrino production with associated photons at a linear collider for
$\l_{1j1} = 0.03$. Solid red (dashed blue) lines correspond to a 
500~GeV (1~TeV)
center-of-mass energy. The cuts of Eq.(\protect\ref{photon_cuts}) 
have been imposed. The points marked with bullets are for a
$\snu_\mu$ resonance at the Snowmass MSugra points 1$a$, 1$b$, 3, 4,
and 5. At the point 2, the sneutrino is beyond the kinematic reach of
the linear collider. If sneutrinos are not distinguished from
anti-sneutrinos, the cross-section(s) would be doubled.}
\protect\label{fig:prodn}
\end{figure}

As the graph shows, we obtain cross-sections typically in the range 50
fb--250 fb.  At a linear collider with around 500 fb$^{-1}$ of integrated
luminosity, this amounts to the production of a very large number of
sneutrinos along with an associated monochromatic photon.  Thus, even if
$\lambda_{1j1}$ were to be smaller by an order of magnitude, we would still
have a fairly large number of such distinctive events.
 It is clear, therefore, that if the sneutrino is kinematically
accessible to a linear collider, low statistics will not be the major
hurdle in their detection.

We focus on the MSugra spectrum and, specifically, on the six representative
points chosen at the 2001 Snowmass conference. The latter along with the 
spectrum are described (for $\mu>0$) in the table below.

\begin{table}[ht]
\begin{center} 
\begin{tabular}{cllll} \\\hline
\bf 1a : & $M_0 = ~100$~GeV, &  $m_{1/2} = 250$~GeV,&  $A_0 = -100$~GeV,
& $\tan\beta = 10$   \\
\bf 1b : & $M_0 = ~200$~GeV, &  $m_{1/2} = 400$~GeV,&  $A_0 = 0$,
&  $\tan\beta = 30$   \\
\bf ~2 : & $M_0 = 1450$~GeV,&  $m_{1/2} = 300$~GeV,&  $A_0 = 0$, 
& $\tan\beta = 10$   \\
\bf ~3 : & $M_0 =  90$~GeV, & $m_{1/2} = 400$~GeV,&  $A_0 = 0$, 
& $\tan\beta = 10$   \\
\bf ~4 : & $M_0 = ~400$~GeV, &  $m_{1/2} = 300$~GeV,&  $A_0 = 0$, 
& $\tan\beta = 50$   \\
\bf ~5 : & $M_0 = ~150$~GeV, &  $m_{1/2} = 300$~GeV,&  $A_0 = -1$~TeV, 
& $\tan\beta =~5$ \\ \hline
\end{tabular}
$$\begin{array}{|c||c|c|c|c|c|c|c|} \hline
{\rm Point} &~\snu_\mu~&~\snu_\tau~&~\stau_1~&
~\stau_2~ &~ \N0_1~ &~ \N0_2~ &~ \Cpm_1~ \\ \hline\hline
1a & 186 & 185 & 133 & 206 & ~96 & 177 & 176 \\
1b & 328 & 317 & 196 & 344 & 160 & 299 & 299 \\
~2 & 1454& 1448& 1439& 1450& ~80 & 135 & 104 \\
~3 & 276 & 275 & 171 & 289 & 161 & 297 & 297 \\
~4 & 441 & 389 & 268 & 415 & 119 & 218 & 218 \\
~5 & 245 & 242 & 181 & 258 & 120 & 226 & 226 \\
\hline
\end{array} $$
\end{center} 
    \label{tbl:spectrum}
\end{table}
Depending on the various decay modes available for the sneutrino, we focus 
on four classes of final states, which are 
(1) $\gamma ee$, (2) $\gamma~ee + E\slash$, (3) $\gamma~\ell_i \ell_j+E\slash$ 
and (4) $\gamma~4\ell + E\slash$. The first kind arises from the direct
$R$-parity-violating decay of the sneutrino and would have a large SM
background from radiative Bhabha scattering. The second and third ones
are obviously reproduced by $WW$-production. The last type arises from
higher-order effects in the SM and has very little background.
This final state arises from the direct $R$-parity
violating decay of the sneutrino into an $e^+e^-$ pair, with, of
course an associated photon from the initial state. The branching
ratio of the sneutrino to this mode is quite significant for
 $\lambda_{1j1} \sim 0.03$ and hence the signal has a reasonable
cross-section. We carry out our analysis at a 500 GeV $e^+e^-$ collider.
To detect this final state, we impose a
set of acceptance sets, namely that each of the particles must not be
too close to the beam pipe,
\beq
\left| \eta (e^\pm) \right| \, , \, \left| \eta (\gamma) \right| < 2.0
      \label{eeg_cuts_rap}
\eeq
and that they should carry  sufficient transverse momenta
\beq
  p_T(e^\pm) > 10 \gev \, \quad {\rm and} \quad
  p_T(\gamma) > 20 \gev \ .
      \label{eeg_cuts_pt}
\eeq
In addition, each pair of the final state particles should be well
separated:
\beq
\delta R > 0.2 \ ,
      \label{eeg_cuts_dr}
\eeq
where $(\delta R)^2 \equiv (\Delta \phi)^2 + (\Delta \eta)^2$ with
$\Delta \eta$  and $\Delta \phi$ respectively
denoting the separation in rapidity and azimuthal angle. On analyzing the 
distributions in various kinematic variables, it is found that an additional
rapidity cut on the difference of the rapidity variable of the final state
electrons ($|\Delta\eta_{ee}|=|\eta_{e^-}-\eta_{e^+}|$) suppresses the huge 
SM background which comes from the t-channel dominated Bhabha scattering 
without reducing the signal by much. A detailed analysis helps us to fix this 
value at 1.7 which we impose for our analysis. The remaining classes of final 
states do not have large SM background and are not affected by the cuts too 
much. So for the leptons and the photon, we choose the cuts to be the same as 
before, namely those listed in Eqs.(\ref{eeg_cuts_rap}--\ref{eeg_cuts_dr}). 
In addition, we demand that the missing transverse momentum be
sufficiently large, {\em viz.}
\beq
       p~\slash_T > 20 \gev
     \label{cut_missing}
\eeq
for it to be considered a genuine physics effect. 

We now focus on the photon in the final state which is our main trigger. We
show the distribution in photon energy for the first two classes of final states
in Fig.~(\ref{fig:gamdist1}) The corresponding fluctuation (Gaussian) in the SM
is also shown at 1, 3 and 5 standard deviations. 
In Fig.~(\ref{fig:gamdist1}a), the clear peaks in the energy distribution of 
the 
\begin{figure}[th] 
\centerline{\psfig{file=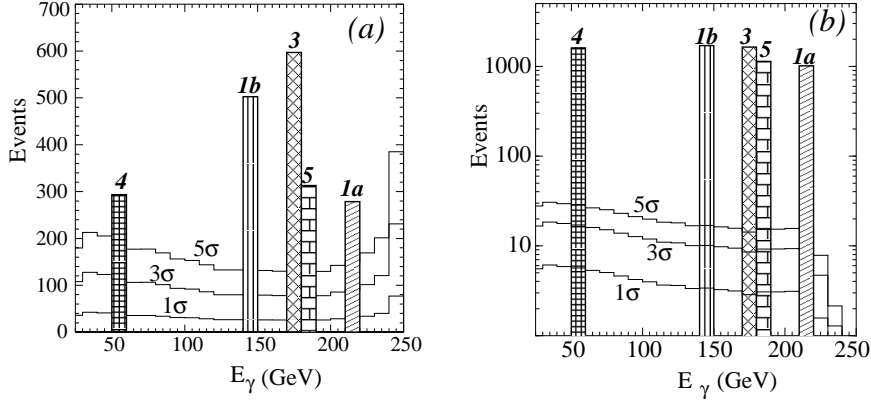,width=4.5in}}
\vspace*{8pt} 
\caption{\it The photon energy distribution when the final states are
(a) $\g e^+e-$ and (b) $\g e^+e^- E\slash$ at a 500 GeV $e^+e^-$ collider. 
The integrated luminosity is $\mathcal{L}=100~fb^{-1}$. The different 
representative Snowmass points are shaded in different patterns.}
 \protect\label{fig:gamdist1}
\end{figure}
photon, over the continuum SM background gives clear hints of the sneutrino 
production. The mass of the sneutrino can also be very easily determined
by the formula given in Eq.~(\ref{eqn:resonance}). However, for class (1), 
we also have the leptons to trigger at and whose invariant mass will reconstruct
the mass of the sneutrino. But the efficacy of the photon signal becomes clear 
when we look at more complicated decays of the sneutrino. In this case the 
final state leptons
cannot reconstruct the sneutrino mass, because of the presence of neutrinos in 
the final state. But the peaks in the photon energy distribution will be a 
complete give-away for the production of sneutrino. This feature gets 
highlighted in Fig.~(\ref{fig:gamdist1}b) where the final state has missing 
energy associated with it. The same feature will repeat itself for the other 
final states. Thus we find that even with complicated decay modes for the 
sneutinos, which render the reconstruction of the parent particles mass quite 
improbable, the clear peaks in the photon energy distribution will save the 
situation. 

\subsection{Extra Dimensions}
Theories with extra dimensions as possible solution to the hierarchy problem
and unification have recently attracted enormous attention and 
interest. We focus on two class of models $viz.$ ADD\cite{ADD} 
and RS\cite{RS} model. In the ADD model 
the Standard Model fields are confined on a (1+3)-dimensional subspace ($D_3$
brane) of a ($1+3+d$)-dimensional spacetime (bulk). The $d$ extra
dimensions are compactified, typically on a $d$-torus of radius $R_c$.
Gravity which reflects the geometry of spacetime itself cannot be confined
and is free to move in the bulk. The essential idea contained in their
model was that when we match the higher dimensional theory with the $4D$
effective theory, the following relation is obtained

\beq
M_{Pl}^2 = M_{S}^{d+2} V_{(d)}= M_{S}^{d+2} (2 \pi R_c)^d  \label{eq:add}
\eeq
where $M_S$ is the fundamental Planck scale (in the bulk), $M_{Pl}$ is
the observed $4D$ Planck scale $\sim 10^{19}$ GeV and $V_{(d)}$ is the 
volume of the extradimensional space. Now if $R_c > 1/M_{Pl}$, then the 
fundamental Planck scale $M_*$ will be lowered from $M_{Pl}$. Under this 
assumption, the fundamental Planck scale can be brought down to values as 
low as $M_S \sim 1$ TeV, consistent with present experimental bounds.
Thus it is able to resolve the hierarchy problem by cutting of the SM
at the TeV scale. The Kaluza Klein (KK) excitations of the (bulk) graviton are
very closed spaced with each mode separated by $1/R_c$ which couple to the SM 
particles by a coupling strength $\sim 1/M_{Pl}$. However there are huge number 
of these excitations which contribute to collectively build up observable 
effects at the electroweak scale. Thus one hopes to see their effect at current 
and future experiments. Another approach is shown in the RS\cite{RS} model, 
which resolves the hierarchy problem, even with small extra dimension. 
They in fact choose a different spacetime geometry (non-factorisable metric) 
and an additional 3-brane in their argument with only one extra dimension 
added to our (1+3)-dimensional spacetime. In contrast to the ADD relation 
Eq.~(\ref{eq:add}), the 4-dimensional Planck scale in the RS approach is:
\beq
M_{Pl}^2 = \frac{M_5^3}{k}\left[1 - e^{-2kR_c\pi}\right] 
\eeq
where $M_5$ is the fundamental scale of the model, $R_c$ the compactification
size and $k$ determines the curvature of the space. A TeV energy scale
can be generated from the 4-dimensional Planck scale if $kR_c\sim 12$,
and thus providing a solution to the hierarchy problem between the
electroweak scale of the standard model and the 4-dimensional Planck
scale. 

The important differences of the RS model with the ADD model are
\begin{itemize}
\item Each KK excitation of the bulk graviton has a mass
\begin{equation}
M_n = x_n {\cal K} e^{-{\cal K}R_c\pi} \equiv x_n m_0
\end{equation}
where $m_0 = {\cal K} e^{-{\cal K}R_c\pi} \sim 100$~GeV is the graviton
mass scale and $x_n$ are the zeros of the Bessel function $J_1(x)$ of
order unity ($n \in \mathbf{Z}$). This means that the Kaluza-Klein
gravitons have masses of a few hundred GeV, unlike the ADD case, where the
masses start from $\sim 1~\mu$eV.
\item Each Kaluza-Klein excitation of the bulk graviton couples to matter
as~\cite{giudice,han,davoudiasl}
\begin{equation}
\kappa e^{{\cal K}R_c\pi} = \frac{4\sqrt{\pi}}{M_P} e^{{\cal K}R_c\pi}
\equiv \frac{4\sqrt{\pi}c_0}{m_0}
\end{equation}
where $\kappa = \sqrt{16\pi G_N}$ and $c_0 = {\cal K}/M_P \simeq 0.01 -
0.1$ is an effective coupling constant, whose magnitude is fixed by ($a$)
naturalness and ($b$) requiring the curvature of the fifth dimension to be
small enough to consider linearized gravity on the `visible' brane.
\end{itemize} 
RS gravitons, thus, resemble weakly-interacting massive particles (WIMPs)
in most models, except for ($a$) the fact that there always exists a tower
of graviton Kaluza-Klein modes and ($b$) these are spin-2 particles. In
phenomenological studies of the RS model, the mass scale $m_0$ and the
ratio $c_0$ may be treated as free parameters\footnote{The alternative
choice of $\Lambda_\pi = \overline{M}_P ~e^{-{\cal K}R_c\pi} =
m_0/\sqrt{8\pi} c_0$ instead of $m_0$ and of ${\cal K}/\overline{M}_P =
\sqrt{8\pi} c_0$ instead of $c_0$ may also be found in the
literature~\cite{davoudiasl}.}: they are convenient replacements for the
fundamental quantities ${\cal K}$ and $R_c$:
\begin{equation}
{\cal K} = c_0 M_P \ , \qquad R_c =
\frac{1}{\pi {\cal K}} \log \frac{{\cal K}}{m_0}
\end{equation}
We show that ``single photon" signals can be very effectively used to look for
such (WIMP) using the radiative return technique. The motivation remains similar
to the exercise done for the sneutrino search. However, a major difference is that
KK gravitons in the RS model follow a distinct relation according to the zeros of
the Bessel function $J_1(x)$ of order unity ($n \in \mathbf{Z}$). Thus, 
by exciting multiple resonances in the photon energy distribution, one could in 
fact check this relation from the appearance of position of the peaks. This 
gives a very strong evidence for RS-type scenario. 

Since the RS gravitons have widths which grow with increasing value of $c_0$, 
they might not be narrow. It can be written down as
\begin{equation}
\Gamma_n = c_0^2 x_n^3 m_0 \sum_P \Delta^{(n)}_{P\bar P}
\label{width}
\end{equation}
where the sum $\sum_P$ runs over all pairs of particles ($G_n \to P\bar
P$)  and the $\Delta^{(n)}_{P\bar P}$ are dimensionless functions of $x_n$
and the ratios $r_P = m_P/m_0$. For details one can look at Ref.\cite{raigrav}.
We calculate the full $2\to3$ process for $e^+e^-\to\g\nu\bar{\nu}$. So here
our final state is just a single hard photon with unbalanced (missing) energy. 
Studies with other mode of decay of the graviton can be found in Ref\cite{mumu}.
We choose the invisible decay mode of the RS gravitons, as this would also 
enable us to compare the signal with the ADD signal, where the equivalent 
process is $e^+e^-\to\g G_n$, and the ADD graviton escapes detection and 
carries the missing energy, because of its very small coupling to SM particles.

We present results for  two values of center-of-mass energy, viz., 
$\sqrt{s} = 1$~TeV and $\sqrt{s} = 2$~TeV. Noting that the final state 
consists of a single hard isolated photon, we impose the following kinematic 
cuts
\begin{itemize}
\item The photon should have energy $E_\gamma \geq 20$~GeV.
\item The photon scattering angle $\theta_\gamma$ should satisfy $15^0
\leq \theta_\gamma \leq 165^0$.
\end{itemize}
These ensure that the tagged photon does not arise from beamstrahlung or
other similar sources. Since the dominant SM background for the 
``single photon" signal comes from the exchange of $W$-bosons in the t-channel,
\begin{figure}[th]
\centerline{\psfig{file=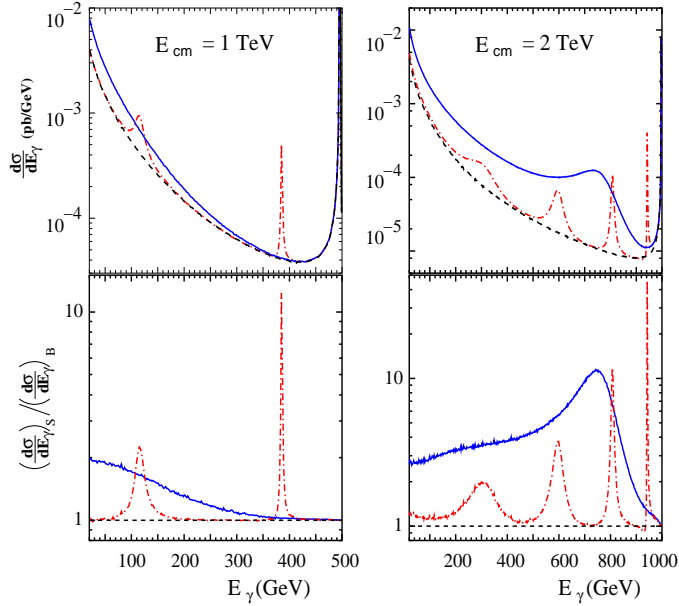,width=3.5in}}
\vspace*{8pt}
\caption{\it 
Energy spectrum of the tagged photon for the
choices $m_0 = 125$~GeV, $c_0 = 0.01$ in red ($-\cdot-$) corresponding to narrow
resonance(s) and $m_0 = 250$~GeV, $c_0 = 0.07$ in blue (solid) corresponding
to broad, indistinct resonances. In the ordinate labels, {\rm S} and
{\rm B} denote signal (SM plus gravitons) and background (SM only)
respectively. Note that the right-most peak (almost flush with the edge
of the box) in the upper graphs, which is due to the
$Z$-boson, can be removed by taking the {\rm S/B} ratio.}
 \protect\label{fig:gravres}
\end{figure}
use of polarized beams (right-polarized electron beam and left-polarized positron
beam), suppresses the background very efficiently and enhances the signal, 
when compared to the case of unpolarized beams. 
We use the partial polarization of the initial beams to be ${\cal P}_e = 0.8$ 
and ${\cal P}_p =-0.6$ for the presented analysis. In Fig.~(\ref{fig:gravres}) 
we show the energy distribution of the single hard photon in the final state. 
The dashed (black) line represents the background (note the $Z$-boson peak
at the extreme right), while the remaining curves correspond to the signal for
low ($-\cdot-$) and high (solid) values of the parameter $c_0$. Sharp resonances
are obtained with the parameter choice $c_0 = 0.01$ and $m_0 = 125$~GeV.
It corresponds to graviton resonances with $M_n \simeq 479, 877, 1272$
and 1665~GeV for $n = 1,2,3$ and 4 respectively. Only the first two are
kinematically accessible at a 1~TeV machine, but all four will be accessible
if the center-of-mass energy rises to 2~TeV. Observe that the resonance peaks 
broaden as the order $n = 1,2, \dots$ of the Kaluza-Klein excitation 
increases\footnote{The present bounds from Tevatron data\cite{tevnu} rule out 
the low lying KK modes ($M_n > 800$ GeV) for the multiple resonances to be seen 
at a 1 TeV machine. However the characteristic features can be still seen at 
the 2 TeV machine or CLIC with 3 TeV center-of-mass energy}.
In the upper halves of the two graphs in Fig.~(\ref{fig:gravres}), we display 
the differential cross-section for the process $e^+e^- \to \gamma + E\slash$. 
The bottom halves show the same distribution, except that now we
exhibit the signal-to-background (S/B) ratio. Not only does this remove
the uninteresting $Z$-peak, but it also takes care of any radiative
corrections, efficiency factors, etc, which can be written in a
factorisable form.

Two (or more) clear resonances seen in the photon spectrum, or rather, in 
the signal-to-background ratio would correspond to a relatively low value of 
$m_0$ and a small value of $c_0$, and would constitute a strong hint of 
RS gravity. For a strong confirmation, positions of the two resonances 
will bear a definite relation if they are due to RS gravitons. 
Using Equation~(\ref{eqn:resonance}) this works out to
the requirement that, for resonance values $E_\gamma^{(1)} > E_\gamma^{(2)}$,
\begin{equation}
\sqrt{\frac{\sqrt{s} - 2E_\gamma^{(1)}}{\sqrt{s} - 2E_\gamma^{(2)}}}
= \frac{M_1}{M_2} = \frac{x_1}{x_2} \simeq 0.546 \ .
\end{equation}
It would be a very remarkable coincidence, indeed, if some other form of 
new physics reproduces such a relation.

A single sharp resonance will allow a consistent fit to the two-parameter 
RS model. One can then look at other distributions, in addition to the single 
photon events to decipher the underlying physics. Just to compare with the case 
of SUSY, a single resonance rules out the conventional $R$-parity conserving 
SUSY. Even with $R$-parity violation, one can use the angular distribution of 
the leptons to differentiate a spin-0 resonance from that of spin-2 graviton 
resonance. 

When the mass $m_0$ and/or the coupling $c_0$ is large, and no
clear resonance structure is discernible, is it possible to distinguish
between the $\gamma + \not{\!\! E}$ signal arising from RS graviton from
those arising from ADD gravitons~\cite{giudice,peskin}? The energy 
distribution of the photon will just show an excess without any bumps 
(which was the give away distinction from the ADD smooth spectrum) in the 
continuous photon energy spectra. However, if one consider this process 
{\it in conjunction} with a benchmark process, like $e^+ e^- \to \mu^+ \mu^-$, 
for example, we do find a marked difference. 
A correlation plot showing the cross-section for the single photon signal 
vis-\'a-vis the cross-section for muon pair-production proves to be clinical
in distinguishing the two different models\cite{raigrav}. 
 
\subsection{Other exotics}

In this section we consider our ``associated photon" signal at a linear
$e^-e^-$ collider to look for doubly charged Higgs bosons which arise in a 
number of physics scenarios\cite{THM1,THM2,LRSM1,LRSM2,LRSM3,LRSM4,LRSM5}, 
the most common models to accommodate such scalars are those with triplet 
Higgs. An added feature often associated with doubly-charged Higgs is the 
possibility of $\Delta L =2$ couplings with leptons which can be very 
effectively probed at $e^-e^-$ colliders and give a strong motivation towards 
running the future linear collider in this mode. The $\Delta L =2$ coupling 
appears in the Lagrangian as 
\begin{equation}
 {L}_Y = i h_{ij}\Psi^T_{iL}C\tau_2\Phi\Psi_{jL} + h.c.
\end{equation}
where $i,j=e,\mu,\tau$ are generation indices, the $\Psi$'s are the 
two-component left-handed lepton fields, and $\Phi$ is the triplet with 
$Y=2$ weak hypercharge. This leads to mass terms for neutrinos once the 
neutral component $\phi^0$ of $\Phi$ acquires a vacuum expectation 
value (vev). Constraints on the $\rho$-parameter puts strong limits
on the the triplet vev translating into 
limits on the L-violation Yukawa couplings which constrain the collider 
signals for doubly-charged scalars sought through $\Delta L = 2$ interactions.  

At linear $e^+e^-$ colliders, such exotics would have to be produced in pair, 
which limits the available phase space for its production. However, they can be 
produced singly if the collision is between electron beams. As before, our 
ignorance of the doubly charged Higgs mass does not allow its production at 
resonance. We point out the usefulness of looking for doubly-charged
scalars  in an $e^- e^-$ collider, in the radiative production channel. 
Not only does this allow us to use the ``radiative return" technique, but also
extends the phase space for its on-shell production, constrained only by 
kinematic cuts on the photon and the machine energy.
With this in view, we consider the process\cite{alan,raidch1,raidch2}
$$
e^{-} e^{-} \longrightarrow \f \gamma \longrightarrow Y \gamma
$$
\noindent
at a $\sqrt{s}=1$ TeV $e^-e^-$ machine, concentrating on the hard 
single photon in the final state. Here $Y$ represents the decay
products of the doubly charged scalar. The hard photon in the final state
will be monochromatic if a doubly-charged resonance is produced, 
irrespective of what it decays into. 
For our analysis, taking the radiative production of the scalar $\f$ 
as the benchmark process, we concentrate only on the flavor diagonal 
coupling $h_{ee}$, which we choose to be $h_{ee}=0.1$ which  respects the 
most stringent bounds coming from muonium-antimuonium conversion 
results which for flavor diagonal coupling is 
$h < 0.44~M_\h~{\rm TeV}^{-1}$ at $90\%$ C.L \cite{Willmann:1998gd}.  
As shown before, the on-shell radiative production of a doubly-charged 
scalar gives an almost monochromatic photon of energy given by 
Eq.~(\ref{eqn:process}).

The major SM background that contributes to the above process is the
radiative Moller scattering process: $e^- + e^- \to \gamma + e^- + e^-$ 
which, although a continuum background, is quite large.
The event selection criteria largely aims at suppressing this continuum
background. We impose the following set of cuts. 
\begin{itemize}
\item Rapidity cut on the final state particles:~
$|\eta(e^-)| < 1.5 ~~~{\rm and}~~~|\eta(\gamma)| < 2.5$
\item minimum cut on energies:
$E(\gamma) > 20 ~{\rm GeV}$, $ E(e^-) > 5 ~{\rm GeV}$
\item to ensure that the final state particles are well separated in
space for the detectors to resolve events : $\delta R > 0.2$
\end{itemize}
Using the above cuts we make an estimate of the SM background and the 
signal. We focus on the main trigger, {\it viz.} the photon. In Fig
\ref{resonance2}(a) we show the distribution of the photon energy, 
where we have
superposed the differential cross-section for {\it signal+background} 
in each bin over the SM background. A pronounced peak can be seen in
the photon energy distribution, due to the monochromaticity of
the photon, corresponding to the recoil energy
against the scalar resonance through the relation of Eq.~(\ref{eqn:process}).
To make our analysis realistic, we have smeared the photon energy by a
Gaussian function whose half-width is guided by the resolution of the
electromagnetic calorimeter \cite{ILC} and also incorporated 
the effects of ISR which often results in substantial broadening of the 
peak. We show the resulting peak for three choices of scalar mass 
(300, 600, 900 GeV). In Fig \ref{resonance2}(a) we only look at the final
\begin{figure}[htbp]
\centerline{\psfig{file=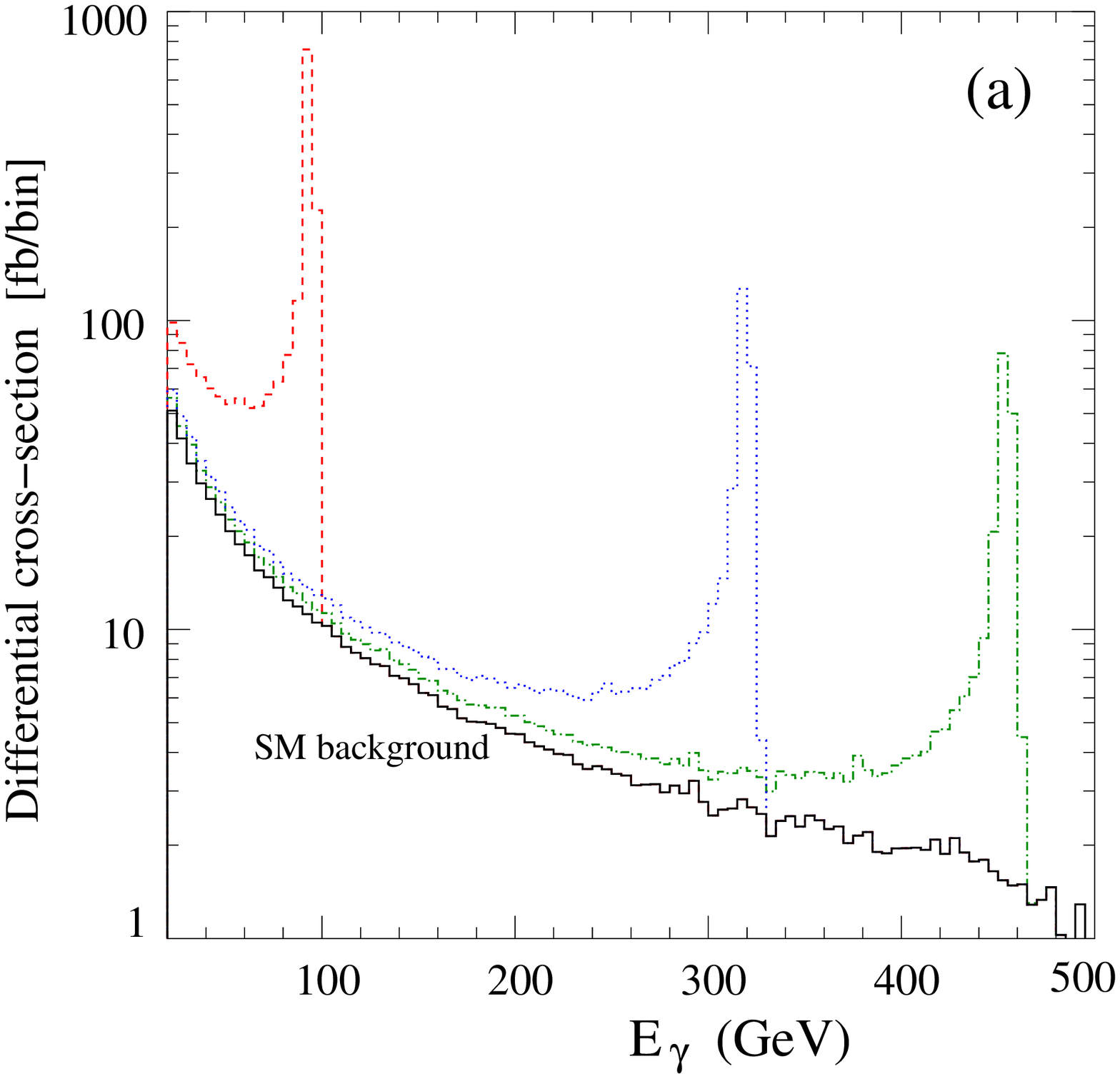,width=2.0in},\psfig{file=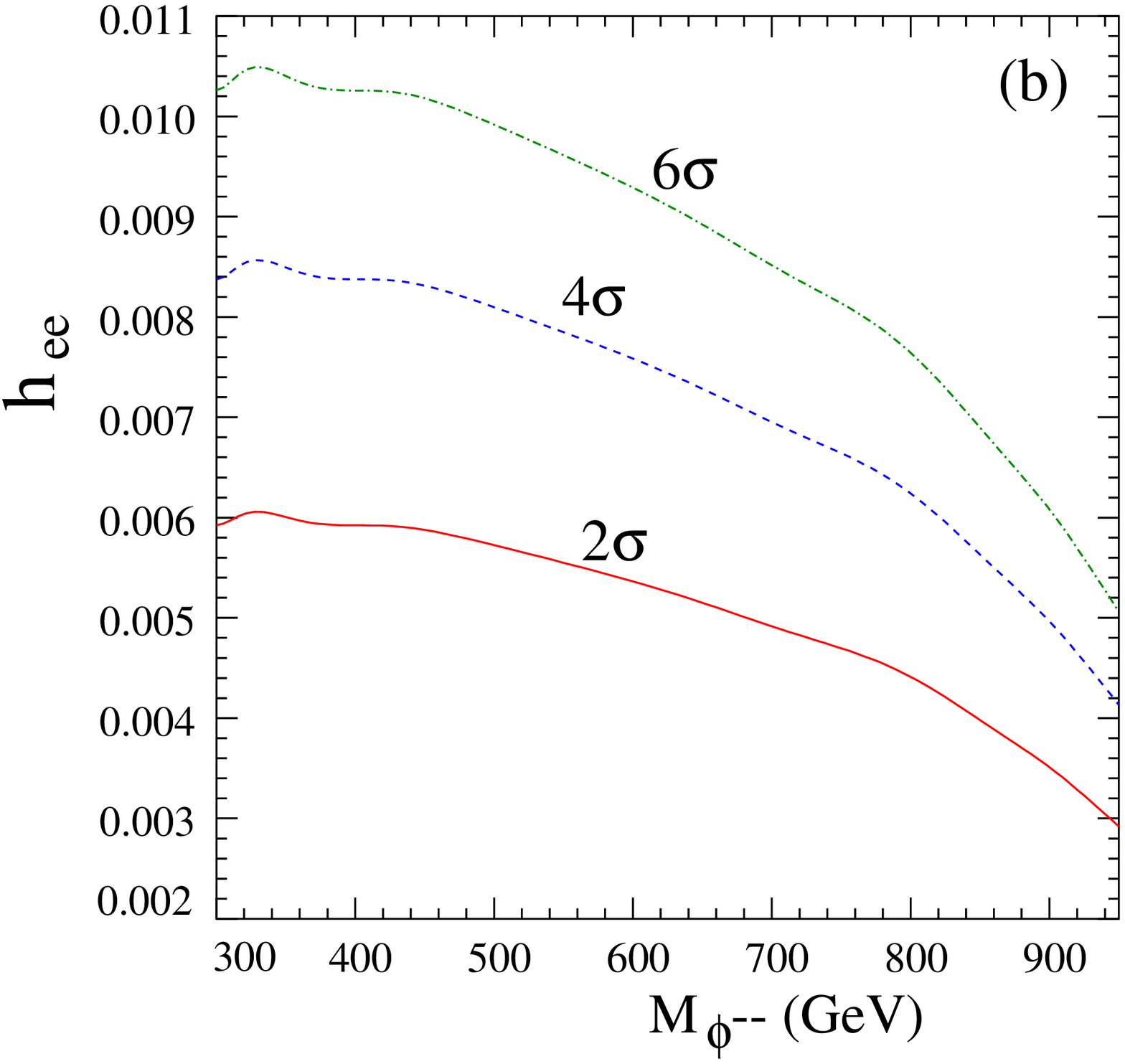,width=2.0in}}
\vspace*{8pt}
\caption{\sl\small {\bf (a)} Differential cross-sections against 
photon energy $E_\gamma$ when $\phi^{--} \to Y$(anything). The 
dash-dot-dash (green) line corresponds to $M_{\phi^{--}} = 300$ GeV, 
dotted (blue) line corresponds to $M_{\phi^{--}} = 600$ GeV and the 
dashed (red) line corresponds to $M_{\phi^{--}} = 900$ GeV respectively.
{\bf (b)} Illustrating the reach of the coupling constant at which the 
resonances in the $E_\gamma$ distribution can be identified over the
fluctuations in the SM background. The assumed luminosity is 
100 fb$^{-1}$.}
\label{resonance2}
\end{figure}
state hard transverse photon in $e^- e^- \to \gamma + \f \to \gamma + Y$(anything).
The distribution again shows peaks corresponding to the
recoil against the massive scalars, irrespective of the knowledge of the
decay products of the scalar. In fact the signal here receives a
relative boost as it is not suppressed by considering any further decay since 
the $BR(\phi^{--}\to Y)= 100\%$. The fact that looking at a single 
photon against the backdrop of a continuum background makes it possible to
identify a LFV ($\Delta L=2$) process in a model independent way, makes
this signal worth studying at a future $e^- e^-$ collider and running
the linear collider in this mode.  
One can also constrain the strength of the $eeH$ coupling. 
Since the rates for the signal depend directly on the $eeH$ coupling 
squared, in Fig \ref{resonance2}(b) we show the strength of the coupling 
for which the peaks would stand out against the fluctuations in the SM 
background. In our analysis we have assumed a luminosity of 
${\mathcal L} = 100~{\rm fb}^{-1}$. The fact that we are not looking at
any specific final state arising from $\f$ decay improves the reach of
this search channel. 
\section{Summary}
\label{sec:concl}
To summarize, the cleanliness of central photon detection at a high energy
linear collider can be very helpful in identifying a weakly interacting massive
particles predicted in various new physics scenarios beyond the SM. To highlight
this, we have considered three different scenarios of physics beyond the SM and
shown how a hard photon in the final state can be useful in extracting the
underlying physics by looking at the photon energy distribution.
The peaks in the hard photon energy can be helpful in two ways. First, 
one does not need to tune the two electron beams, and can therefore work 
without a prior knowledge of the particles ($X$) mass against which the photon 
recoils. Secondly, this method is shown to work even if the $X$ dominantly 
decays into states that are not clean enough for the resonance to be identified.
Thus, as soon as one succeeds in reducing the SM backgrounds, one can clearly 
see non-SM interactions, just by looking at the accompanying hard photon. 
More exotic resonances such as extra $Z^ \prime$ bosons predicted in models 
beyond SM with extra U(1)'s, KK gauge bosons predicted in the universal extra 
dimension models (UED), bileptons, etc., are amenable to detection in this 
manner. Although we have focussed primarily on resonant searches, the 
associated photon signals are also sensitive to new physics 
signals\cite{godfrey,choi,sinphot1,sinphot2} in other forms.

Another way of using associated photons to capture resonances at future 
linear colliders would be through beamstrahlung and ISR photons, by looking at 
the simple scattering process $ee\to X^*\to PP$ where $P$ is any SM particle. 
The radiated photons will cause a spread in the beam energy and cause some of 
the events to take place at values lower than the actual center-of-mass 
energies ($\sqrt{s}$). If the exchange particle mass is less than $\sqrt{s}$
and the events due to radiation happen around the resonance(s), it will
provide a huge enhancement in the cross-section\cite{rairad1,rairad2}. Thus,
associated photon signals will play an important role at the next generation 
linear colliders and the radiative return technique proves to be a crucial tool,
both in the case of tagged (hard) photons or radiative photons which go down 
the beam pipe, but may cause large energy spread to excite new resonances 
in the invariant mass distributions of the final state. 
The physics of ``associated/single photon" signals at the next generation of 
linear colliders merits a lot of attention and can prove to be an important 
tool to study physics beyond the SM.

\bigskip
\noindent {\large\bf Acknowledgments}
The author would especially like to thank S. Raychaudhuri for his invaluable 
guidance and for introducing him to the idea of associated photon signals.
The author also acknowledges all his 
collaborators, especially R.M. Godbole, B. Mukhopadhyaya and D. Choudhury for 
their contributions towards his understanding of the subject. 
The author also thanks M. Perelstein, K.S. Babu, J.L. Feng,
T. Han, K. Matchev, S. Nandi, T. Rizzo and M. Peskin for their support and 
useful comments on the topic during his visit to the US. The author gratefully
acknowledges support from the Academy of Finland (project number 115032).   



\begin{thebibliography}{0}

\bibitem{neuexp}
Y.~Fukuda {\it et al.}  [Super-Kamiokande Collaboration],
Phys.\ Rev.\ Lett.\  {\bf 81}, 1562 (1998)
[arXiv:hep-ex/9807003].

\bibitem{higgsM}
P.~W.~Higgs,
Phys.\ Lett.\  {\bf 12}, 132 (1964).

\bibitem{ILC}
  R.~D.~Heuer,
  Nucl.\ Phys.\ Proc.\ Suppl.\  {\bf 154}, 131 (2006).

\bibitem{history}
  M.~S.~Chen and P.~M.~Zerwas,
  Phys.\ Rev.\  D {\bf 11}, 58 (1975).
\bibitem{Fayet}
  P.~Fayet,
  Phys.\ Lett.\  B {\bf 117}, 460 (1982).
\bibitem{Ellis}
  J.~R.~Ellis and J.~S.~Hagelin,
  Phys.\ Lett.\  B {\bf 122}, 303 (1983).
\bibitem{Grifols}
  J.~A.~Grifols, M.~Martinez and J.~Sola,
  Nucl.\ Phys.\  B {\bf 268}, 151 (1986).

\bibitem{mont1}
  G.~Montagna, M.~Moretti, O.~Nicrosini and F.~Piccinini,
  Nucl.\ Phys.\  B {\bf 541}, 31 (1999)
  [arXiv:hep-ph/9807465].

\bibitem{mont2}
  G.~Montagna, M.~Moretti, O.~Nicrosini, M.~Osmo and F.~Piccinini,
  Eur.\ Phys.\ J.\  C {\bf 21}, 291 (2001)
  [arXiv:hep-ph/0103155].

\bibitem{anomalous1}
  A.~Denner, S.~Dittmaier, M.~Roth and D.~Wackeroth,
  Eur.\ Phys.\ J.\  C {\bf 20}, 201 (2001)
  [arXiv:hep-ph/0104057].
\bibitem{anomalous2}
  P.~Achard {\it et al.}  [L3 Collaboration],
  Phys.\ Lett.\  B {\bf 527}, 29 (2002)
  [arXiv:hep-ex/0111029].

\bibitem{Ma}
  E.~Ma and J.~Okada,
  Phys.\ Rev.\ Lett.\  {\bf 41}, 287 (1978)
  [Erratum-ibid.\  {\bf 41}, 1759 (1978)].

\bibitem{PDG}
  W.~M.~Yao {\it et al.}  [Particle Data Group],
  J.\ Phys.\ G {\bf 33}, 1 (2006).

\bibitem{godfrey}
  S.~Godfrey, P.~Kalyniak, B.~Kamal and A.~Leike,
  Phys.\ Rev.\  D {\bf 61}, 113009 (2000)
  [arXiv:hep-ph/0001074].
\bibitem{choi}
  S.~Y.~Choi, J.~S.~Shim, H.~S.~Song, J.~Song and C.~Yu,
  Phys.\ Rev.\  D {\bf 60}, 013007 (1999)
  [arXiv:hep-ph/9901368].


\bibitem{sinphot1}
  A.~Datta, A.~Datta and S.~Raychaudhuri,
  Phys.\ Lett.\  B {\bf 349}, 113 (1995)
  [arXiv:hep-ph/9411435].
\bibitem{sinphot2}
  C.~H.~Chen, M.~Drees and J.~F.~Gunion,
  Phys.\ Rev.\ Lett.\  {\bf 76}, 2002 (1996)
  [arXiv:hep-ph/9512230].
\bibitem{sinphot3}
  A.~Datta, A.~Datta and S.~Raychaudhuri,
  Eur.\ Phys.\ J.\  C {\bf 1}, 375 (1998)
  [arXiv:hep-ph/9605432].
\bibitem{sinphot4}
  A.~Datta and A.~Datta,
  Phys.\ Lett.\  B {\bf 578}, 165 (2004)
  [arXiv:hep-ph/0210218].
\bibitem{raigrav}
  S.~Kumar Rai and S.~Raychaudhuri,
  JHEP {\bf 0310}, 020 (2003)
  [arXiv:hep-ph/0307096].
\bibitem{sinphot5}
  H.~K.~Dreiner, O.~Kittel and U.~Langenfeld,
  Phys.\ Rev.\  D {\bf 74}, 115010 (2006)
  [arXiv:hep-ph/0610020].

\bibitem{RPV}
  H.~K.~Dreiner,
  arXiv:hep-ph/9707435.

\bibitem{rplimits}
  R.~Barbier {\it et al.},
  Phys.\ Rept.\  {\bf 420}, 1 (2005)
  [arXiv:hep-ph/0406039].

\bibitem{raisneutrino}
  D.~Choudhury, S.~K.~Rai and S.~Raychaudhuri,
  Phys.\ Rev.\  D {\bf 71}, 095009 (2005)
  [arXiv:hep-ph/0412411].

\bibitem{ADD}
  N.~Arkani-Hamed, S.~Dimopoulos and G.~R.~Dvali,
  Phys.\ Lett.\  B {\bf 429}, 263 (1998)
  [arXiv:hep-ph/9803315].

\bibitem{RS}
  L.~Randall and R.~Sundrum,
  Phys.\ Rev.\ Lett.\  {\bf 83}, 3370 (1999)
  [arXiv:hep-ph/9905221].

\bibitem{giudice}
  G.~F.~Giudice, R.~Rattazzi and J.~D.~Wells,
  Nucl.\ Phys.\  B {\bf 544}, 3 (1999)
  [arXiv:hep-ph/9811291].

\bibitem{han}
  T.~Han, J.~D.~Lykken and R.~J.~Zhang,
  Phys.\ Rev.\  D {\bf 59}, 105006 (1999)
  [arXiv:hep-ph/9811350].

\bibitem{davoudiasl}
  H.~Davoudiasl, J.~L.~Hewett and T.~G.~Rizzo,
  Phys.\ Rev.\ Lett.\  {\bf 84}, 2080 (2000)
  [arXiv:hep-ph/9909255].

\bibitem{mumu}
  T.~Buanes, E.~W.~Dvergsnes and P.~Osland,
  Eur.\ Phys.\ J.\  C {\bf 35}, 555 (2004)
  [arXiv:hep-ph/0403267].

\bibitem{tevnu}
  V.~M.~Abazov {\it et al.}  [D0 Collaboration],
  arXiv:0710.3338 [hep-ex].

\bibitem{peskin}            
  E.~A.~Mirabelli, M.~Perelstein and M.~E.~Peskin,
  Phys.\ Rev.\ Lett.\  {\bf 82}, 2236 (1999)
  [arXiv:hep-ph/9811337].

\bibitem{THM1}
  J.~Schechter and J.~W.~F.~Valle,
  Phys.\ Rev.\ D {\bf 23}, 1666 (1981).
\bibitem{THM2}
  G.~B.~Gelmini and M.~Roncadelli,
  Phys.\ Lett.\ B {\bf 99}, 411 (1981).

\bibitem{LRSM1}
  J.~C.~Pati and A.~Salam,
  Phys.\ Rev.\ D {\bf 10}, 275 (1974).
\bibitem{LRSM2}
  R.~N.~Mohapatra and J.~C.~Pati,
  Phys.\ Rev.\ D {\bf 11}, 566 (1975).
\bibitem{LRSM3}
  G.~Senjanovic and R.~N.~Mohapatra,
  Phys.\ Rev.\ D {\bf 12}, 1502 (1975).
\bibitem{LRSM4}
  R.~N.~Mohapatra and D.~P.~Sidhu,
  Phys.\ Rev.\ Lett.\  {\bf 38}, 667 (1977).
\bibitem{LRSM5}
  R.~E.~Marshak and R.~N.~Mohapatra,
  Phys.\ Lett.\ B {\bf 91}, 222 (1980).

\bibitem{alan}
  R.~A.~Alanakian,
  Phys.\ Lett.\ B {\bf 402}, 130 (1997)
  [arXiv:hep-ph/9605217].

\bibitem{raidch1}
  B.~Mukhopadhyaya and S.~K.~Rai,
  Phys.\ Lett.\ B {\bf 633}, 519 (2006)
  [arXiv:hep-ph/0508290].

\bibitem{raidch2}
  B.~Mukhopadhyaya and S.~K.~Rai,
  Pramana {\bf 69}, 809 (2007)
  [arXiv:hep-ph/0608112].

\bibitem{Willmann:1998gd}
  L.~Willmann {\it et al.},
  Phys.\ Rev.\ Lett.\  {\bf 82} (1999) 49
  [arXiv:hep-ex/9807011];

\bibitem{rairad1}
  N.~K.~Mondal {\it et al.},
  Pramana {\bf 63}, 1331 (2004)
  [arXiv:hep-ph/0410340].

\bibitem{rairad2}
  R.~M.~Godbole, S.~K.~Rai and S.~Raychaudhuri,
  Eur.\ Phys.\ J.\  C {\bf 50}, 979 (2007)
  [arXiv:hep-ph/0611156].
\end{thebibliography}
\end{document}